\documentclass[referee]{raa}
\usepackage{graphicx,times}
\usepackage{natbib}
\usepackage{amssymb,amsmath}
\bibpunct{(}{)}{;}{a}{}{,}

\usepackage[a4paper=true,dvipdfm=true,pagebackref=true]{hyperref}
\hypersetup{pdftitle = The title of my PDF, pdfauthor = My name, pdfsubject= The subject, pdfkeywords = keyword1 keyword2 keyword3}
\hypersetup{colorlinks = true, linkcolor = green, anchorcolor = red, citecolor = blue, filecolor = red, pagecolor = red, urlcolor = red}

\begin{document}

   \title{The first photometric investigation of the contact binary OQ Cam
}

 \volnopage{ {\bf 20XX} Vol.\ {\bf X} No. {\bf XX}, 000--000}
   \setcounter{page}{1}

   \author{Ya-Ni Guo\inst{1}, Kai Li\inst{1,2}, Qi-Qi Xia\inst{1}, Xing Gao
      \inst{2},  Lin-Qiao Jiang\inst{3}, Yuan Liu\inst{4}
   }

   \institute{ Shandong Key Laboratory of Optical Astronomy and Solar-Terrestrial Environment, School of Space Science and Physics, Institute of Space Sciences, Shandong University, Weihai, Shandong, 264209, China; {\it kaili@sdu.edu.cn}\\
        \and
             Xinjiang Astronomical Observatory, 150 Science 1-Street, Urumqi 830011, China\\
	\and
Artificial intelligence Key Laboratory of Sichuan Province, Sichuan University of Science \& Engineering, Zigong 643000, China\\
\and
Qilu Institute of Technology, Jinan 250200, China\\
\vs \no
   {\small Received 20XX Month Day; accepted 20XX Month Day}
}

\abstract{The first charge-coupled device photometric light curves in B, V, R${_c}$, and I${_c}$ bands of the short-period contact binary star OQ Cam are presented. Through analyzing the light curves with the Wilson-Devinney code, it is found that OQ Cam is a W-type shallow contact binary star with a contact degree of f =20.2\%. Based on the O-C analyzing, the orbit period displays a long term increasing with a rate of $dP/dt=4.40\times10{^{-7}} day \cdot year{^{-1}}$. The increasing in orbit period can be interpreted by mass transfer from the less massive component to the more massive one. As the orbit period increasing, OQ Cam would evolve from the present shallow contact configuration to a none contact stage. So it may be a potential candidate to confirm the thermal relaxation oscillation theory.
}

   \authorrunning{Guo et al. }            
   \titlerunning{The first photometric investigation of OQ Cam}  
   \maketitle

%
\section{Introduction}           
\label{sect:intro}

In the Milky Way, almost half of the stars are binary stars. EW-type binaries's main characteristics are that primary minimum is almost as deep as the secondary minimum of the light curve, their orbital periods are shorter than one day and two components have mass transfer frequently \citep{He}. Normally, EW-type binary stars are contact binary systems, which refer to both two components filling their Roche Lobes. The two component stars have a common convection envelope, located between the inner and outer critical surfaces of the Roche model. They have mass transfer frequently through inner Lagrangian point. And the two components have close surface temperatures \citep{Qian,Qian2017}. The mass transfer has a great influence on the evolution of the close binary systems. It would change the mass distributions of binary systems, further leading to variations of the orbital periods. There are several other reasons for the long-term variations of orbital period: Mass loss from the binary system, which makes the period decreasing; Angular momentum loss (AML) from the binary system via a magnetic stellar wind or by the third body removing the angular momentum from the central system, which permanently makes the period decreasing; Common convective envelope (CCE) dominated mechanism. This mechanism is especially applied to high mass ratios. The variation of orbit period depends on the fill-out factor that is regarded as a measurement of the thickness of CCE \citep{Liu,Zhao}. The third body has a significant effect on the evolution of binary system. Generally speaking, the presence of a third body will lead to stellar mergers, X-ray binaries, gamma-ray bursts and so on \citep{Zhao}.
Through the investigation of eclipsing binary stars, many parameters can be determined for example the orbit period variation, mass, temperature, radius, luminosity and so on. According to a large number of observations, there is a popular belief that contact binary stars are mainly formed from detached or semi-detached binary stars with cool components, and they will evolve into rapidly rotating single stars \citep{Roxburgh,Bradstreet1994,Stepien}. The evolution of contact binary is relying on the angular momentum left in the system, mass transfer rate and angular momentum loss rate. Some typical contact binaries'evolutions are presented by \citet{Stepien,Stepien1}. However, this conclusion is still under debate at present and further research is still needed, so we have to study more contact binaries.

Thermal relaxation oscillation theory \citep{Lucy,Flannery,Robertson} is an important model for the structure and evolution of marginal contact binary stars. This theory includes two phases: during contact phase, mass is transferred from the less massive component to the more massive component, during the non-contact phase, mass is returned to the less massive component. With the increasing of the period, the shallow contact binary stars may evolve from contact phase into non-contact phase. Contact binary stars are divided into W-type systems and A-type ones. If the more massive component is hotter, the binary star is an A-type binary star. If the more massive component is cooler, the binary star is a W-type binary star \citep{Binnendijk,Binnendijk1977}. Research has shown that when $q\geq0.4$, there are more W-type binaries than A-type; $0.25\leq q \leq0.4$, there are a even distribution of W-type binaries and A-type ones; $q\textless0.25$, the number of A-type binaries are more than W-type binaries \citep{Kim}. \citet{Zhang2020} has presented the secondary stars of A-type binaries are evolved from initial more massive stars, while the ones of W-type are formed via mass transfer. Some scientists believe that W-type binaries will evolve into A-type ones \citep{Hilditch,Awadalla,Eker,Gazeas} and someone believes that it is no evolutionary relationship between two types \citep{Yildiz}. This argument is under debate.

The eclipsing star OQ Cam was firstly discovered by \citet{Khruslov2006} through the data of Northern Sky Variability Survey (NSVS, \cite{Wozniak2004}). \citet{Khruslov2006} has given that the light curve of this binary star is EW-type and the orbit period is 0.43784 days. At present, the photometric solution and the orbit period investigation of OQ Cam have not been published. In this paper, we analyzed the four-color light curves and orbital period variation of OQ Cam for the first time.

\section{CCD Photometric Observation}
\label{sect:Obs}

We have observed OQ Cam using 60cm Ningbo Bureau of Education and Xinjiang Observatory Telescope (NEXT) and the Weihai Observatory 1.0-m telescope of Shandong University (WHOT) \citep{Hu et al. 2014} from 2018 to 2019. The CCDs of NEXT and WHOT both have $2K\times2K$ pixels. And their field of views are $22'\times22'$ and $12'\times12'$, respectively. The standard Johnson-Cousin-Bessel B,V,R${_c}$,I${_c}$ filters have been used. The details of observation are listed in Table 1. The variable, comparison and check star are shown in Figure 1, and some of their information is displayed in Table 2.

\begin{figure}
\centering
\includegraphics[width=8cm, angle=0]{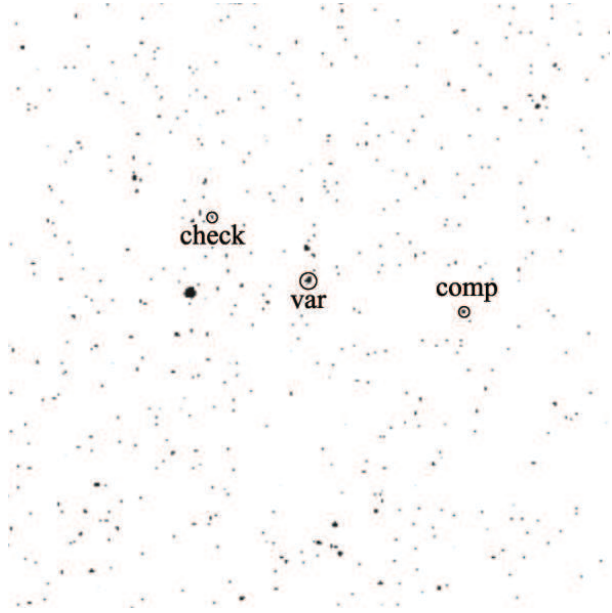}
\caption{The image of OQ Cam in B band observed by NEXT. The "var" refers to the variable star, the "comp" refers to the comparison star and "check" refers to the check star.}
\label{Figure1}
\end{figure}

We used the C-Munipack \nolinebreak\footnotemark[1] \footnotetext[1]{http://c-munipack.sourceforge.net/} program to process the CCD data. The bias and flat corrections have been done for all images. The aperture photometry and differential photometry were used to reduce the CCD data. Then, the magnitude differences between the variable and comparison stars, and those between the comparison and check stars were determined. The four-color light curves observed on 26, December, 2018 are plotted in Figure 2. The x-axis is the Heliocentric Julian Date (HJD), and the y-axis is the magnitude difference. From the light curves, we can find that the light curves change continuously and the secondary minima are as deep as the primary minima. And the magnitude differences between the comparison and check stars are plain, suggesting that both the comparison and check stars are not variable stars.

\begin{figure}
\centering
\includegraphics[width=8cm, angle=0]{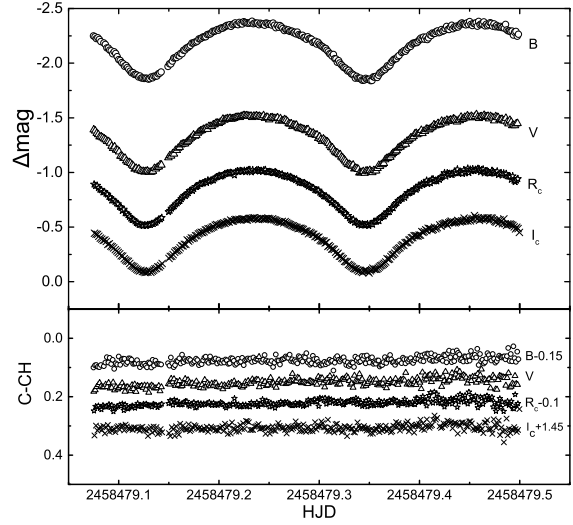}
\caption{The figure displays the light curves of OQ Cam observed by NEXT on December 26, 2018. The bottom of this figure is the magnitude difference between the comparison and check stars. "C" and "CH" refer to the comparison and check stars, respectively.}
\label{Figure2}
\end{figure}

\begin{table}
\bc
\begin{minipage}[]{100mm}
\caption[]{observation information\label{tab1}}\end{minipage}
\setlength{\tabcolsep}{1pt}
\small
 \begin{tabular}{ccccccccccccc}
  \hline\noalign{\smallskip}
 Date     & Filter and exposure time&type         &Telescope \\
\hline
 2018.12.24& B60sV40sR$_c$50sI$_c$40s&minimum &NEXT      \\
 2018.12.25& B30sV20sR$_c$20sI$_c$20s&minimum &NEXT      \\
 2018.12.26& B20sV5sR$_c$5sI$_c$5s&light curve&NEXT      \\
 2019.11.01&V10s                    &minimum &WHOT      \\
  \noalign{\smallskip}\hline
\end{tabular}
\ec
\end{table}

\begin{table}
\bc
\begin{minipage}[]{100mm}
\caption[]{Details of the Variable, Comparison and Check Stars\label{tab2}}\end{minipage}
\setlength{\tabcolsep}{1pt}
\small
 \begin{tabular}{ccccccccccccc}
  \hline\noalign{\smallskip}
 Star     & Name   &    R.A.    &    Decl.  & J(mag) & H(mag) & K (mag)\\
\hline
 Variable star  & OQ Cam        &03$^h$33$^m$34.32$^s$ &+64$^{\circ}$16$^{'}$46$^{''}$.2 & 9.638 & 9.351 & 9.292     \\
 Comparison star& 2MASS J03324359+6415410 &03$^h$32$^m$43.59$^s$ &+64$^{\circ}$15$^{'}$41$^{''}$.0 & 9.362 &8.725  &8.503     \\
 Check Star& 2MASS 03340661+6418584   &03$^h$34$^m$06.61$^s$ &+64$^{\circ}$18$^{'}$58$^{''}$.4 & 8.663 &7.833  &7.605    \\
  \noalign{\smallskip}\hline
\end{tabular}
\ec
\end{table}
\section{Photometric solutions}

According to the complete light curves of OQ Cam, the secondary minima are almost as deep as the primary minimum, indicating OQ Cam is a typical EW-type binary star. The Wilson-Devinney (W-D) program \citep{aau07,aau08,aau09} was applied to model the light curves. Gaia DR 2 (\citealt{aau05,aau06}) has determined a surface temperature of 5676K for OQ Cam, which is treated as the temperature of the primary star. This surface temperature is estimated through the approach of Priam algorithm according to the Gaia three-band photometry with an accuracy of about 324 K in the range 3000 K to 10 000 K \citep{Rene}. The gravity-darkening coefficients and the bolometric albedos were setting as $g_{1,2}$= 0.32 and $A_{1,2}$ = 0.5 due to the the convective envelopes \citep{aau04}. The bolometric and bandpass limb-darkening coefficients were chosen from \citet{Van Hamme}'s table.

No investigation has been carried out for OQ Cam, hence we used the method of q-search to determine its mass ratio. The details are as follows: the mass ratio was fixed at a range from 0.1 to 10, and the step length is 0.1. Then we can obtain a series of convergent solutions. In the processing of running the program, orbital inclination, the temperature of secondary star, the luminosity of the primary star in BVR${_c}$I${_c}$ bands, and the dimensionless potential of the primary star are adjustable parameters. The sum square of residuals versus mass ratio is shown in Figure 3 (in this figure, we only showed the range from 0.2 to 2.5). From the figure, we can find when the mass ratio is 0.4, the sum square of residuals is the smallest. So this value was set as the original mass ratio and an unfixed value. By running the W-D code, the final results were obtained and are shown in Table 3. All the errors in Table 3 are not real ones but only fitting errors computed by the WD program, which are underestimated \citep{Prasad}. If one wants to obtain the real errors, Bayesian modeling with Markov Chain Monte Carlo (MCMC) should be applied to model the light curves of eclipsing binaries. The fitting curve is shown in the top panel of Figure 4, and the O-C residuals are shown in the bottom panel. Very flat O-C residuals indicate that the theoretical light curves fit the observed light curves very well. The light curve is symmetric, no spot was needed. In addition, we have searched for the third light, but the value is always negative, so no third light was detected.

\begin{table}
\bc
\caption[]{Photometric results\label{tab1}}\
\setlength{\tabcolsep}{1pt}
\small
 \begin{tabular}{ccc}
  \hline\noalign{\smallskip}
Parameters &                   Value&                    Errors      \\
\hline
     $g_1=g_2$ &               0.32 &                   Assumed     \\
     $A_1=A_2$ &               0.5  &                   Assumed     \\
     $T_1(K)$  &               5676 &                   Assumed     \\
     $T_2(K)$  &               5721 &                   6     \\
     $q(M_2/M_1)$ &            0.479&                 0.007     \\
     $\Omega_{in}$ &           2.835&                    --        \\
     $\Omega_{out}$ &          2.548&                     --        \\
     $\Omega_1$=$\Omega_2$ &   2.777&                 0.014     \\
     $i$ &                     73.1&                0.1     \\
     $(L_1/{L_1+L_2})_B$&         0.647&                 0.001     \\
     $(L_1/{L_1+L_2})_V$&         0.650&                 0.001     \\
     $(L_1/{L_1+L_2})_R{_c}$&         0.652&                 0.001     \\
     $(L_1/{L_1+L_2})_I{_c}$&         0.653&                 0.001     \\
     $r_1(pole)$ &             0.428&                 0.001     \\
     $r_1(side)$ &             0.457&                 0.002     \\
     $r_1(back)$ &             0.489&                 0.002     \\
     $r_2(pole)$ &             0.307&                 0.005     \\
     $r_2(side)$ &             0.322&                 0.007     \\
     $r_2(back)$ &             0.363&                 0.012     \\
     $f$&                      20.2\%&                  4.9\%     \\
  \noalign{\smallskip}\hline
\end{tabular}
\ec
\end{table}

\begin{figure}
\centering
\includegraphics[width=8cm, angle=0]{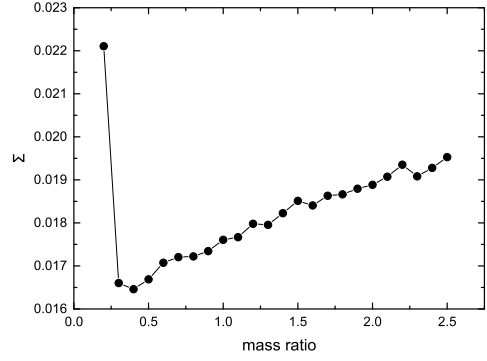}
\caption{This figure displays the sum square of residuals $\Sigma$ versus mass ratio q of OQ Cam.}
\label{Figure3}
\end{figure}

\begin{figure}
\centering
\includegraphics[width=8cm, angle=0]{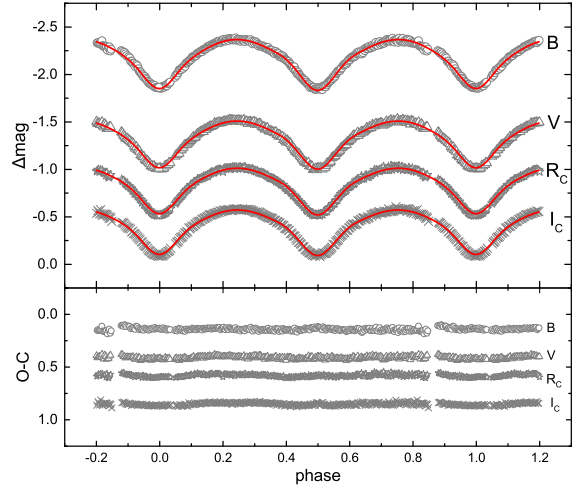}
\caption{This figure displays the observed light curves and fitting curves in B,V,R${_c}$, I${_c}$ bands of OQ Cam, and the values of O-C are shown in the bottom.}
\label{Figure4}
\end{figure}

\section{Orbital period investigations}
\label{sect:Obs}
A total of 29 minimum moments have been collected from the literatures and our observation data. In which, 22 minimum moments were observed by CCD, and 7 were observed by PE. The time range of these 29 minimum moments is from 1999 to 2019. Based on these minima, we analyzed the orbital period change of OQ Cam. All the minimum moments are listed in Table 4. At first, the O-C values were calculated using the following linear ephemeris:
\begin{equation}
\label{E1}
Min(HJD)=2458479.12807+0.{^d}43784\times E.
\end{equation}
Then, a linear correction was applied to Equation (1), and a revised linear ephemeris was determined as follows:
\begin{equation}
\label{E1}
Min(HJD)=245879.12127(\pm0.00182)+0.4378401(\pm0.0000004)\times E.
\end{equation}
Using this ephemeris, the final O-C values were calculated.
The relationship between epoch and O-C is shown in Figure 5. Then we used the least square method to fit the values of O-C and determined the following equation:
\begin{eqnarray}
\label{E2}
Min(HJD)&=&2458479.12706(\pm0.00048)+0.4378435(\pm0.0000002)\times E\\\nonumber
&+&2.64(\pm0.15)\times 10{^{-10}}\times E{^2}.
\end{eqnarray}
The corresponding fitting curve labeled by black line is displayed in Figure 5. According to Equation (3), the orbital period exhibits a long term increasing, and the rate is $dP/dt=4.40(\pm0.25)\times10{^{-7}} day \cdot year{^{-1}}$.

\begin{table}
\bc
\begin{minipage}[]{100mm}
\caption[]{The minimum moments of OQ Cam\label{mbh}}\end{minipage}
\setlength{\tabcolsep}{2.5pt}
\small
 \begin{tabular}{ccccccccccccc}
  \hline\noalign{\smallskip}
 JD (Hel.)   &  Error  &   Method   &  Min.&  epoch  & 	O-C      & Residuals    & Ref.     \\
2400000+     &         &            &      &         &           &              &          \\ \hline
51501.7180   &  -      &    CCD     &   p  &-15936.0 &-0.0166    &  -0.0014     &   (1)     \\
53225.7017   &  0.0010 &    CCD     &   s  &-11998.5 &-0.0049    &  0.0024      &   (2)     \\
55609.2946   &  0.0002 &    PE      &   s  & -6554.5 &-0.0037    &  0.0017      &   (3)     \\
55932.6372   &  0.0003 &    CCD     &   p  & -5816.0 &-0.0061    &  -0.0008     &   (4)     \\
56187.4611   &  0.0001 &    PE      &   p  & -5234.0 &-0.0051    &  -0.0001     &   (5)     \\
56205.8503   &  0.0001 &    CCD     &   p  & -5192.0 &-0.0052    &  -0.0002     &   (6)     \\
56227.9635   &  0.0005 &    CCD     &   s  & -5141.5 &-0.0029    &  0.0020      &   (7)     \\
56246.3500   &  0.0001 &    CCD     &   s  & -5099.5 &-0.0057    &  -0.0008     &   (8)     \\
56371.3541   &  0.0005 &    CCD     &   p  & -4814.0 &-0.0049    &  -0.0003     &   (9)     \\
56585.8951   &  0.0010 &    CCD     &   p  & -4324.0 &-0.0056    &  -0.0014     &   (10)     \\
57196.4674   &  0.0005 &    CCD     &   s  & -2929.5 &-0.0013    &  0.0007      &   (11)     \\
57270.4614   &  0.0001 &    PE      &   s  & -2760.5 &-0.0023    &  -0.0006     &   (12)     \\
57296.2957   &  0.0010 &    PE      &   s  & -2701.5 &-0.0005    &  0.0010      &   (12)     \\
57296.5118   &  0.0006 &    PE      &   p  & -2701.0 &-0.0034    &  -0.0018     &   (12)     \\
57298.4832   &  0.0041 &    PE      &   s  & -2696.5 &-0.0022    &  -0.0007     &   (12)     \\
57327.3804   &  0.0003 &    PE      &   s  & -2630.5 &-0.0025    &  -0.0011     &   (12)     \\
57595.5565   &  0.0002 &    CCD     &   p  & -2018.0 &-0.0035    &  -0.0034     &   (11)     \\
58090.7602   &  0.0002 &    CCD     &   p  & -887.0  &0.0031    &  0.0002      &   (13)     \\
58117.4729   &  0.0002 &    CCD     &   p  & -826.0  &0.0076    &  0.0044      &   (14)     \\
58136.5151   &  0.0002 &    CCD     &   s  & -782.5  &0.0038    &  0.0005      &   (14)     \\
58465.5538   &  0.0002 &    CCD     &   p  & -31.0   &0.0056    &  -0.0001     &   (14)     \\
58477.3727   &  0.0023 &    CCD     &   p  & -4.0    &0.0028    &  -0.0030     &   (15)     \\
58478.2524   &  0.0023 &    CCD     &   p  & -2.0    &0.0068    &  0.0011      &   (15)     \\
58479.1281   &  0.0003 &    CCD     &   p  & 0.0     &0.0068     &  0.0010      &   (15)     \\
58479.3463   &  0.0003 &    CCD     &   s  & 0.5     &0.0061    &  0.0003      &   (15)     \\
58534.2954   &  0.0003 &    CCD     &   p  & 126.0   &0.0043    &  -0.0020     &   (14)     \\
58576.3278   &  0.0004 &    CCD     &   p  & 222.0   &0.0060     &  -0.0006     &   (14)     \\
58789.1207   &  0.0001 &    CCD     &   p  & 708.0   &0.0086     &  -0.0003     &   (15)     \\
58821.3043   &  0.0002 &    CCD     &   s  & 781.5   &0.0110     &  0.0024      &   (14)     \\
  \noalign{\smallskip}\hline
\end{tabular}
\ec
\tablecomments{0.86\textwidth}{(1)\cite{Khruslov2006}; (2)This paper (WASP); (3)\cite{HUBSCHER1}; (4)\cite{DIETHELM ROGER}; (5)\cite{HUBSCHER2}; (6)\cite{NELSON}; (7)\cite{DIETHELM1}; (8)\cite{HON}; (9)\cite{Corfini}; (10)\cite{DIETHELM2}; (11)\cite{Jur}; (12)\cite{Hubscher3}; (13)\cite{Nelson}; (14)http://var2.astro.cz/brno/protokoly.php; (15)This paper.}
\end{table}

\begin{figure}
\centering
\includegraphics[width=8cm, angle=0]{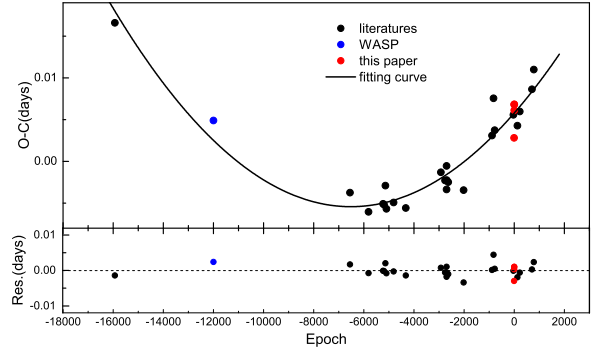}
\caption{This figure displays the O-C values of OQ Cam, the black line respects the fitting curve of Equation (2). The residuals are shown in the bottom.}
\label{Figure5}
\end{figure}

\section{Discussion}
\label{sect:discussion}
According to the results of the photometric solutions, OQ Cam is proven to be a shallow contact binary star with a contact degree of 20.2\% and a mass ratio of 0.479 . Moreover, the temperature difference between two components is 45K, indicating the binary is in thermal contact. By analyzing all minima, it is found that the orbital period manifests a long term increasing with a rate of $4.40\times10{^{-7}} day \cdot year{^{-1}}$. Like II UMa \citep{Zhou2016}, NO Cam \citep{Zhou2017}, TYC 1337-1137-1 \citep{Liao2017}, this can be explained by mass transfer from the less massive component to the more massive one.
According to the temperature of the primary star, its corresponding spectral type is G5 \citep{Cox}. Assuming the primary star is a main-sequence star and based on \citet{Cox}, the mass of the primary star can be estimated to be $M{_1}=0.92M_{\odot}$. The mass of the secondary star is calculated: $M{_2}=M{_1}\times 0.473=0.44M_{\odot}$. Using the following equation:
\begin{eqnarray}
\label{E2}
\frac{\mathrm d M_2}{\mathrm d t}=\frac{M_1M_2}{3P(M_1-M_2)}\times\frac{\mathrm dP}{\mathrm dt},
\end{eqnarray}
the mass transfer rate was calculated to be $\mathrm d M_2/\mathrm d t=2.82\times 10^{-7}M_{\odot}/year$.

We also calculated the distance between the two component stars based on their masses and gained $a=2.69R_{\odot}$ according to $M_1+M_2=0.0134a^3/P^2$. The radii of the primary and secondary stars can be computed to be $R_1=(r_{1pole}{\cdot}r_{1side}{\cdot}r_{1back})^{1/3}{\cdot}a=1.23R_{\odot}$ and $R_2=(r_{2pole}{\cdot}r_{2side}{\cdot}r_{2back})^{1/3}{\cdot}a=0.89R_{\odot}$, respectively. The luminosities of the primary and secondary stars are $L_1=(T_1/T_\odot)^4\cdot(R_1/R_{\odot})^2=1.19L_{\odot}$ and $L_2=(T_2/T_\odot)^4\cdot(R_2/R_{\odot})^2=0.79L_{\odot}$\, respectively. To intuitively reflect the evolution of the two stars, we drew the mass-luminosity (M-L) diagram as shown Figure 6. "ZAMS" and "TAMS" refer to zero-age main sequence and terminal-age main sequence, respectively, which are constructed based on the binary evolution code provided by \citet{Hurley}. From the M-L diagram, the primary star is located between the ZAMS and TAMS, meaning that it is a main sequence star, while the secondary star is located above the TAMS, indicating that the secondary star is at an advanced evolutionary stage. The evolutionary stages of the two components are similar with those of other W-type contact binaries. Table 4 shows some W-type shallow contact binaries with increasing period. As seen in Table 5, OQ Cam has a similar orbital period increasing rate to the other stars.

OQ Cam shows a long term increasing in orbit period and its mass ratio is 0.479. This is agree with the result that for the W-type contact stars with $q\textgreater0.4$, their periods show a long term increasing, while $q\textless0.4$, their periods show a long term decreasing \citep{Qian2001}. Due to the increasing period, the distance between the primary and secondary stars will be increasing. Further, OQ Cam may evolve from the present shallow contact stage to the none-contact stage. This is consistent with the thermal relaxation oscillation theory \citep{Lucy,Flannery,Robertson}. So it may be a potential candidate to confirm the thermal relaxation oscillation theory. Further observations and investigations are needed.

\begin{figure}
\centering
\includegraphics[width=8cm, angle=0]{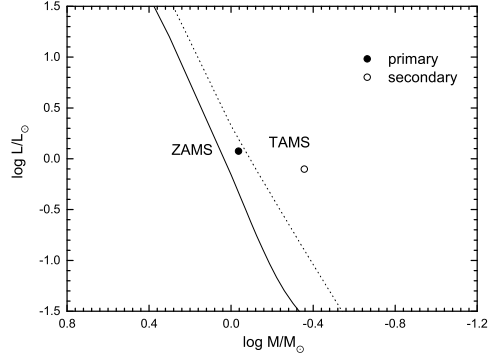}
\caption{This figure displays M-L diagram. The solid circle refers to the primary star
and hollow circle refers to the secondary star. The solid and dashed lines represent ZAMS and TAMS, respectively.}
\label{Figure6}
\end{figure}

\begin{table}
\bc
\begin{minipage}[]{100mm}
\caption[]{W-subtype Shallow Contact Binaries with Increasing Orbital Period\label{mbh}}\end{minipage}
\setlength{\tabcolsep}{2.5pt}
\small
 \begin{tabular}{ccccccccccccc}
  \hline\noalign{\smallskip}
star	 &  period$(day)$    &  q	  &$f(\%)$ &T${_1}(K)$  &	T${_2}(K)$	   & $dp/dt (day\cdot yr{^{-1}}) $     &	reference\\ \hline
AR Boo	 &  0.344874        & 0.382  &12.3	   &5100	         &5398	        &$2.68\times 10{^{-7}}$                &\cite{Lee}\\
AD Cnc   &	0.282729        & 0.770  &8.3	   &4790	         &5000	        &$4.94 \times 10{^{-7}}$                &\cite{Qian2007}\\
FZ Ori   &  0.399986  	     & 0.860  &2       &5940	         &5983	        &$2.28 \times 10{^{-8}}$               &\cite{Prasad}\\
RZ Com	 &  0.338507        & 0.425  &20.1    &4900	         &5000	        &$3.97 \times 10{^{-8}}$               &\cite{He1}\\
FP Lyn   &	0.359085        & 0.867	  &13.4	   &5683	         &5909	        &$4.19 \times 10{^{-7}} $	           &\cite{Michel}\\
FV CVn	 &  0.315365        & 0.930	  &4.6     &5120	         &5470	        &$7.70 \times 10{^{-7}} $	           &\cite{Michel}\\
PS Vir   &  0.289806        & 0.305  &14.4	   &5800	         &5976	        &$1.26 \times 10{^{-7}}$               &\cite{Yuan}\\
TY UMa	 &  0.354548        & 0.396  &13.4	   &6229             &6250	        &$5.18\times10{^{-7}}$                 & \cite{Li}\\
J140533.33+114639.1  &0.225123& 0.645 &7.9     &4563             &4680          &$2.09\times10{^{-7}}$                 &\cite{Zhang1}\\
V509 Cam &  0.350347       &0.364    &10      &5662             &5902          &$3.96 \times 10{^{-8}}$                &\cite{Li1}\\
AE Phe   &  0.362375       &0.392    &14.6    &5940             &5983          &$6.17 \times 10{^{-8}}$                &\cite{He2}\\
  \noalign{\smallskip}\hline
\end{tabular}
\ec
\tablecomments{0.86\textwidth}{(1) q=M${_2}$/M${_1}$; (2) The subscript 1, 2 represent the massive and less massive star, respectively.}
\end{table}

\normalem
\begin{acknowledgements}

This work is supported by National Natural Science Foundation of China (NSFC) (No. 11703016), and by the Joint Research Fund in Astronomy (Nos. U1631108 and U1931103) under cooperative agreement between NSFC and Chinese Academy of Sciences (CAS), and by the Natural Science Foundation of Shandong Province (Nos. ZR2014AQ019, ZR2017PA009, ZR2017PA010, JQ201702), and by Young Scholars Program of Shandong University, Weihai (Nos. 20820162003, 20820171006), and by the Open Research Program of Key Laboratory for the Structure and Evolution of Celestial Objects (No. OP201704). RM acknowledge the financial support from the UNAM under DGAPA grant PAPIIT IN 100918. Thanks Dr. Lohr very much for sending us eclipsing times and uncertainties of the targets observed by SuperWASP. This work is partly supported by the Supercomputing Center of Shandong University, Weihai.

This work has made use of data from the European Space Agency (ESA) mission
{\it Gaia} (https://www.cosmos.esa.int/gaia), processed by the {\it Gaia}
Data Processing and Analysis Consortium (DPAC,
https://www.cosmos.esa.int/web/gaia/dpac/consortium). Funding for the DPAC
has been provided by national institutions, in particular the institutions
participating in the {\it Gaia} Multilateral Agreement.

This paper makes use of data from the DR1 of the WASP data \citep{Butters2010} as provided by the WASP consortium,
and the computing and storage facilities at the CERIT Scientific Cloud, reg. no. CZ.1.05/3.2.00/08.0144
which is operated by Masaryk University, Czech Republic.
\end{acknowledgements}

\label{lastpage}
\end{document}